# Ultrathin metasurface with topological transition for manipulating spoof surface plasmon polaritons

Yihao Yang[1,2,3,4], Lian Shen[1,2,3], Liqiao Jing[1,2,3], Zheping Shao[1,2,3], Thomas Koschny[4], Costas M. Soukoulis[4], Hongsheng Chen[1,2,3*]

[1]*State Key Laboratory of Modern Optical Instrumentation, Zhejiang University, Hangzhou 310027, China.*

[2]*College of Information Science & Electronic Engineering, Zhejiang University, Hangzhou 310027, China.*

[3]*The Electromagnetics Academy at Zhejiang University, Zhejiang University, Hangzhou 310027, China.*

[4]*Department of Physics and Astronomy and Ames Laboratory-U.S. DOE, Iowa State University, Ames, Iowa 50011, USA.*

\* *hansomchen@zju.edu.cn.*

## Abstract

Metasurfaces, with intrinsically planar nature and subwavelength thickness, provide us unconventional methodologies to not only mold the flow of propagating waves but also manipulate near-field waves. Plasmonic metasurfaces with topological transition for controlling surface plasmon polaritons (SPPs) recently have been experimentally demonstrated, which, however, are limited to optical frequencies. In this work, we proposed and experimentally characterized an ultrathin metasurface with the topological transition for manipulating spoof SPPs at low frequency. We demonstrated rich interesting phenomena based on this metasurface, including frequency-dependent spatial localization, non-diffraction propagation, negative refraction, and dispersion-dependent spin-momentum locking of spoof SPPs. Comparing with traditional three-dimensional metamaterials, our metasurface exhibits low propagation loss and compatibility with the photonic integrated circuit, which may find plenty of applications in spatial multiplexers, focusing and imaging devices, planar hyperlens, and dispersion-dependent directional couplers, in microwave and terahertz frequencies.

# Introduction

Metasurfaces are with planar profile and subwavelength thickness composed of arrays of optical scatterers[1]. The metasurfaces introduce abrupt changes of optical properties, which is due to the strong interaction between light and the subwavelength scatterers or antennas, and can control the amplitude[2], phase[3], and polarization[4] of propagating light at will. The fascinating properties of planar metasurfaces provide us unparalleled methodologies to manipulate the manner of propagating light and promise a bright future for useful devices with compact volume[5,6]. The metasurfaces can not only mold flow of the propagating light but also manipulate near-field waves, such as surface plasmon polaritons (SPPs)[7]. The SPPs propagate at the interface between metasurface and the dielectric materials, and by tailoring the dispersion of the metasurface we can manipulate the propagation and spin manners of the SPPs. Based on this concept, a metasurface composed of silver/air grating was theoretically proposed, whose dispersion gradually changes from elliptical to flat and finally to hyperbolic, showing an unprecedented capability to control the SPPs[7]. This metasurface with feasible geometries has been experimentally realized recently by applying lithographic and etching techniques[8]. Comparing with the traditional bulky metamaterials, the planar metasurfaces have lower propagation loss and compatibility with the integrated metamaterials circuit. It shows plenty of interesting phenomena, including negative refraction of SPPs propagation, non-diffraction SPPs propagation, and dispersion-dependent plasmonic spin Hall effect. It may find broad applications in planar focusing and imaging devices, hyperlens, integrated photonic circuits, and quantum optics.

The metasurfaces with the topological transition (MTT) will be extremely useful not only at optical spectrums, but also at the low frequencies, *i.e.*, far-infrared, terahertz, and microwave frequencies. Though successfully demonstrated, this optical metasurface can't be directly applied at low frequencies, as the noble metals, *e.g.*, silver and gold, behave akin to perfect electric conductor (PEC) at low frequencies, and the surface wave modes known as Sommerfeld or Zenneck waves[9] with weak confinement are difficultly controlled by using the metal/air grating structure. In the far-infrared frequency, the graphene can support SPPs due to the strong light-matter interaction. Gomez-Diaz *et al.* have proposed an ultrathin metasurface based on graphene strips[10]. These uniform graphene sheets can be treated as a homogenous anisotropic conductive surface and exhibit topological transition of equifrequency contours (EFCs) associated with a dramatic tailoring of the local density of electromagnetic states. However, realizing such a graphene-based metasurface will need to face some technical challenges, and it has not been realized yet.

In 2004, the spoof SPPs was proposed by Pendry *et al.*, which are composed of structured PECs to mimic the optical properties of SPPs, *e.g.*, dispersion behavior and light confinement[11]. Though physically different, *i.e.*, the SPPs arise from the interaction between light and the free electrons in noble metals, while the spoof SPPs result from the interaction between electromagnetic wave and the spatial capacitances and inductances induced by the structured metal surface. The intrinsically close connections between SPPs and spoof SPPs have brought a plenty of phenomena, such as localized SPPs[12-14], and applications in SPPs, *e.g.*, sensor[15], laser beams[16], imaging, and directive emission[17], to spoof SPPs. This also gives us a clue to design an MTT operational at low frequency, namely, an MTT for manipulating spoof SPPs.

Therefore, in this paper, we proposed and experimentally demonstrated an MTT for manipulating the spoof SPPs. Based on an equivalent model theory, we design a complementary H-shape resonator metasurface (CHRM), whose EFCs in the wave vector space is experienced a topological transition from a closed elliptical curve to a straight line, and finally to an open hyperbolic curve. As a result, normal and non-divergent diffraction, negative refraction are achieved when the spoof SPPs propagate along the metasurface. This metasurface also shows a dispersion-dependent spin-momentum locking of spoof SPPs, where anomalous spin Hall effect will be shown when the EFCs are hyperbolic. Our metasurface will find potential applications in the spatial multiplexer, focusing and imaging devices, hyperlens, dispersion-dependent directional coupler, and photonic integrated circuits.

## Theories

The unit cell of the ultrathin MTT for manipulating spoof SPPs is shown in Fig.1 (a), which is a sandwich structure: copper ground layer, substrate with permittivity of 2.55, and a complementary H-shape layer[18]. The electromagnetic properties of this unit cell are controlled by the geometries, for example, $p$=6 mm, $l$=5 mm, $w$=0.5 mm, $g$=0.25 mm, $t$=1 mm, and the thickness of the metal is 0.035 mm in our case. When imposing EM waves to the metasurface, surface currents will be induced and oscillate on the metal surface due to the special geometries. It is obvious that due to the lack of C4 symmetry, this unit cell design will show different responses when EM wave propagates along *x* and *y* directions, respectively. For better understanding the unit cell design, we give equivalent circuit models when spoof SPPs propagating along *x* and *y* directions in Fig. 1 (b) and (c), respectively. Thus, we can get the spoof surface plasma frequency, $f_x = 1/\sqrt{L(C_2 C_3 + C_1(C_2 + C_3)/(C_1 + C_3))}$ when wave vector is along an *x* direction; $f_y = 1/\sqrt{L'(C_2'/2 + C_1')}$ when wave vector is along the *y*-direction, which mimics the surface plasmon frequency of the noble metals in optical frequency. When $f_0 < f_x < f_y$ (in this case, $f_x < f_y$, $f_0$ is the operational frequency), the spoof SPPs can propagate along both orthogonal directions, the EFC is elliptical. Whereas when $f_x < f_0 < f_y$, only the spoof SPPs along y direction exists, by properly altering the geometries, it may tailor the EFCs to be hyperbolic.

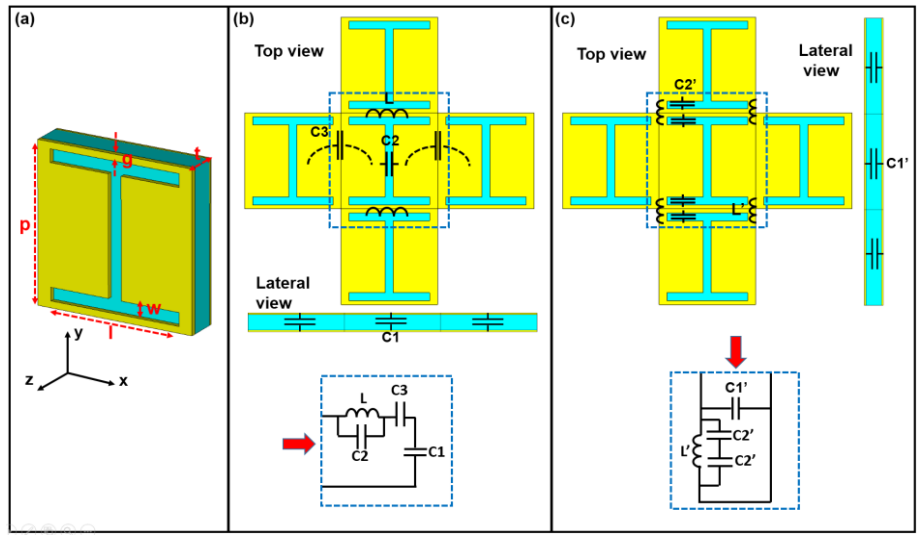

**Figure 1.** (a) Unit cell of an ultrathin MTT, where *p*= 6 mm, *l*=5 mm, *w*=0.5 mm, *g*=0.25 mm, *t*=1 mm, and the thickness of the copper layer is 0.035 mm. (b)-(c) Equivalent circuit models when EM waves propagating along x and y directions, respectively.

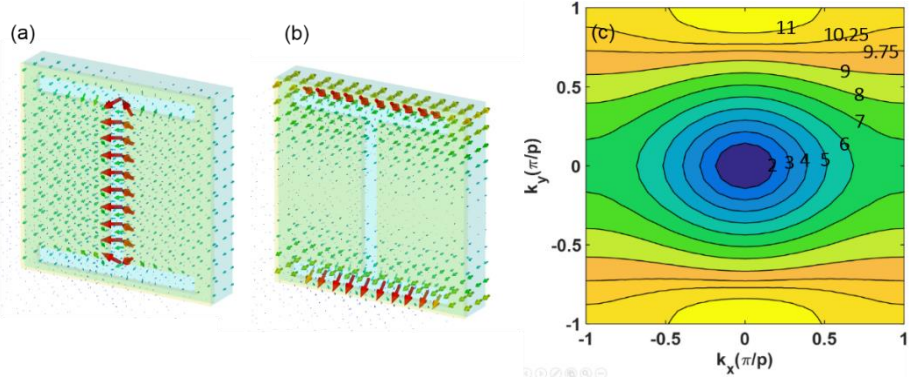

**Figure 2.** (a)-(b) Electric field distributions when EM waves propagating along x and y-direction, respectively. (c) The first-band EFCs in the first Brillouin zone. Frequency values are in units of $\pi/p$.

We numerically calculate the dispersion of the metasurface by employing the eigenvalue module of commercial software CST (Computer Simulation Technology) Microwave Studio. Electric field distributions when EM waves propagating along x and y directions are shown in Fig. 2(a) and (b), respectively. The different field distributions indicate that the metasurface manipulates the spoof SPPs in different manners. Besides, electric field distributions also manifest that the gap-induced conductances play the main roles in the equivalent circuit models. The first-band EFCs in the first Brillouin zone of the CHRM unit cell is shown in Fig. 2(c), from which, we can see a topology transition from a closed elliptical dispersion (below 6.0 GHz), to an open extreme anisotropic one and flat one (9.0 GHz to 10.0 GHz), and finally to an open hyperbolic one (10.0 GHz to 11.0 GHz). It manifests that through controlling geometry-induced capacitances and inductances, we can tailor the dispersion of the metasurface to manipulate the propagation manner of spoof SPPs on the metasurface. Note that this unit cell will show a topological transition of EFCs in wave vector space with only the top metal layer, however, considering the excitation of spoof SPPs in experiment, we add a ground layer. Moreover, the additional ground layer gives more freedoms to control and excite the spoof SPPs, such as the directional coupler we designed at the last of this paper.

## Simulations and experiments

In the following, we will demonstrate the main electromagnetic properties of the CHRM, including the topological transition of EFCs in wave vector space, frequency-dependent spatial localization, non-diffraction propagation, negative refraction, and dispersion-dependent spin-momentum locking of spoof SPPs. In Fig.3, we show the electric field distributions when exciting the surface wave mode with an electric dipole between two metal layers of the metasurface. The size of metasurface here is 300 mm*384 mm, and the simulation is performed in the transient module of CST Microwave Studio. In the Figures, we show the components of the electric field along z direction at 5.6 mm over the metasurface and the corresponding EFCs at 7.5 GHz, 8.75 GHz, 9.25 GHz, 9.75 GHz, 10.25 GHz, and 10.75 GHz, respectively. From the figures, one can see that the wavefronts of the spoof SPPs on the *XY* plane is gradually changed from convex, to flat, and to concave. At 6.0 GHz, the corresponding EFC is a closed ellipse and the spoof SPPs can propagate along any direction on the *XY* plane (Fig. 3(a)). When the frequency increases to 8.75 GHz, the EFC becomes extreme anisotropic. As the group velocity vector of the spoof SPPs should be perpendicular to the EFCs, the spoof SPPs are guided and split into two beams as a consequence of the special shapes of EFCs. The propagation direction of the spoof SPPs is frequency-dependent, as shown from Fig. 3(b) and (c). This phenomenon can be applied to design a spatial multiplexer. At the transition point around 9.75 GHz, the spoof SPPs propagates with self-collimation manner due to the flat dispersion[19]. This self-collimation phenomenon has been also found in photonic crystals and maybe find potential applications in the integrated surface wave circuit system and hyperlens[20] (Fig. 3(d)). From 9.75 GHz to 11 GHz, the spoof SPPs propagate with convergent manners due to the hyperbolic EFCs (Fig. 3(e) and (f)).

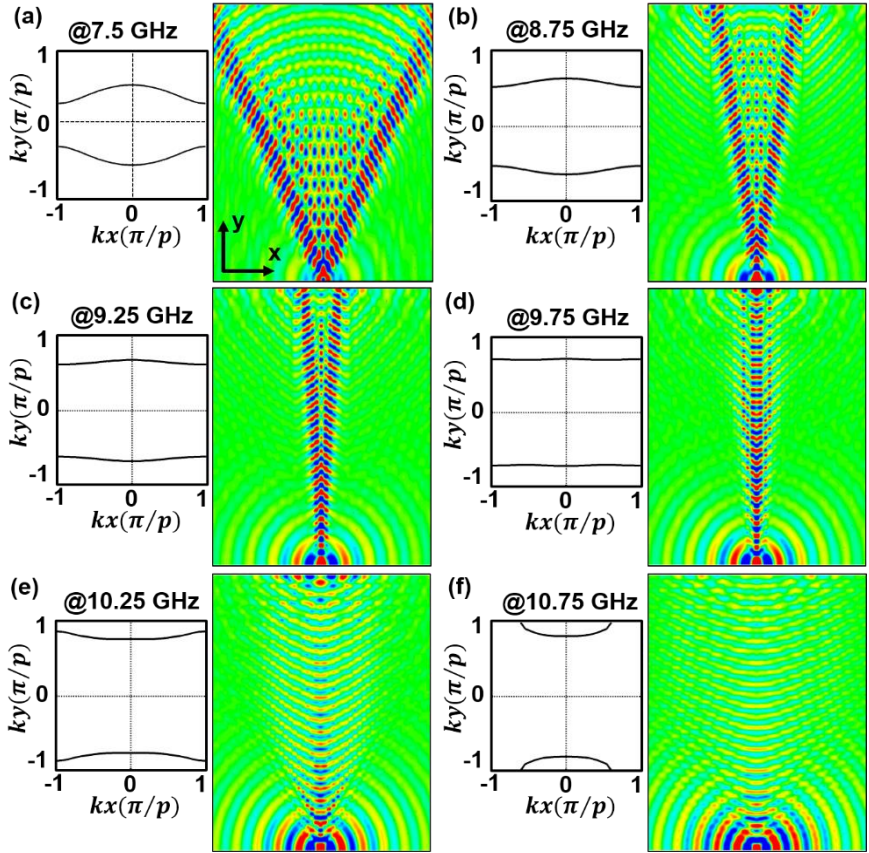

**Figure 3.** EFCs and simulated $E_z$ field distributions on *XY* plane 5.6 mm over the metasurface at 7.5GHz (a), 8.75 GHz (b), 9.25 GHz (c), 9.75 GHz (d), 10.25 GHz (e), and 10.75 GHz (f).

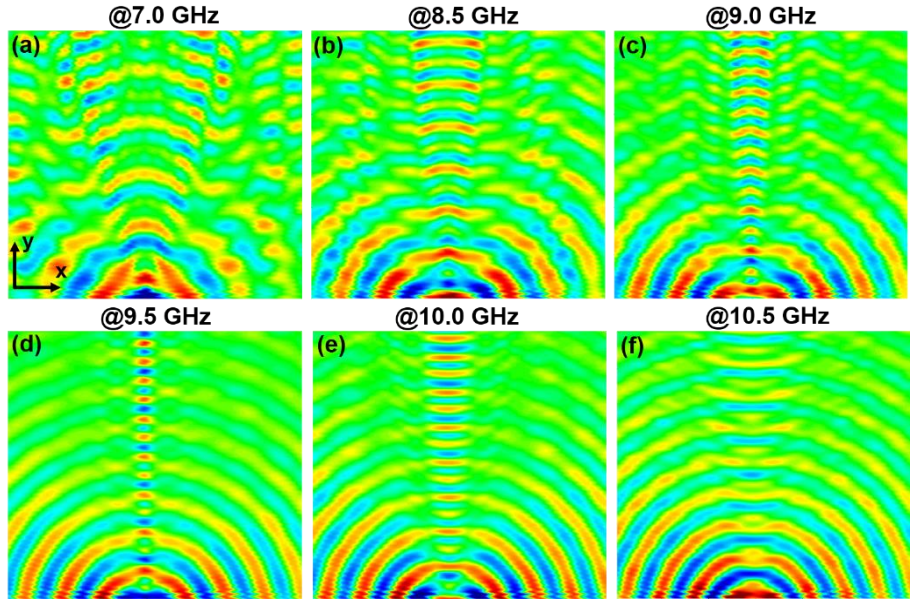

**Figure 4.** Measured $E_z$ field distributions on *XY* plane 1 mm over the metasurface at 7.0 GHz (a), 8.5 GHz (b), 9.0 GHz (c), 9.5 GHz (d), 10.0 GHz (e), and 10.5 GHz (f).

This topological transition phenomenon has been also experimentally demonstrated. In the implementation, we printed the complementary H-shape resonator structure on a substrate, which is a 1 mm Teflon woven glass fabric copper-clad laminates with a permittivity of 2.55 and $\tan(\delta) < 0.001$ at 10.0 GHz. We used a dipole between two metal layers to excite the spoof SPPs, the same source setting in the simulations. Therefore, the measured field distributions will be exactly matched with the simulated ones as shown in Fig. 4. In order to obtain the field distributions, we used a dipole antenna 1 mm over the metasurface to detect the z-oriented electric ($Ez$) field point to point by a three-dimensional movement platform, and the measured region is 240 mm * 230 mm. From Fig. 4, we can directly observe the transition of wavefronts from a convex, to flat, and to concave. The transition point shifts slightly from 9.75 GHz to 9.5 GHz, due to the imperfection of the fabrication. At the transition point, the spoof SPPs propagate with non-diffraction manner, which is very similar to the spatial solitons in nonlinear optics, however, it's only based on a linear optical system[21].

If properly designing the dispersion of the background metasurface (Fig. 5(b)), negative refraction of spoof SPPs will occur at the interference between CHRM and background medium, as shown in Fig. 5(a). The EFCs of the background metasurface and the CHRM at 10.6 GHz is shown in Fig. 5(c). One can see that all incoming wave vectors at such a frequency are included within the EFCs of the CHRM, thus, all-angle negative refraction is enabled, which can be applied to surface waves focusing and imaging[22]. In the experiment, we chose air as the background, when the spoof SPPs scatter into the air, it will be focused, as shown in Fig. 5(d). Therefore, the CHRM can work as an ultrathin planar imaging device, which will be very useful, especially at terahertz and far-infrared frequencies.

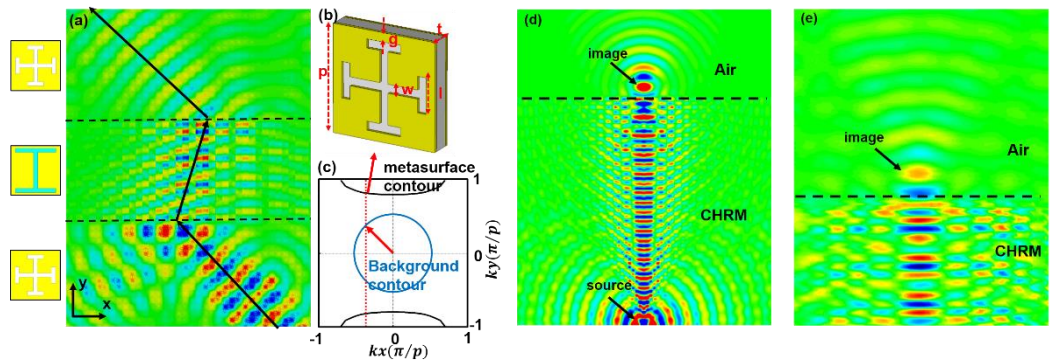

**Fig. 5. Negative refraction of spoof SPPs, and focusing and imaging devices.** (a) *Ez* field distribution at an *XY* plane 0.5 mm below the patterned metal surface. When spoof SPPs on a background metasurface are incident in the hyperbolic metasurface, negative refraction phenomena will occur at the interface between them at 10.6 GHz. (b) Geometry parameters of the background metasurface, where *p*=5 mm, *g*=0.5 mm, *w*=0.5 mm, *l*=2 mm, *t*=1 mm, and the material between two metal layers is air. (c) Constant-frequency dispersion contour at 10.6 GHz. The blue and the black curves are the dispersion contours for background metasurface and hyperbolic metasurface, respectively. The red arrow represents the propagation direction of spoof SPPs. (d)-(e) Simulated and measured *Ez* field on an *XY* plane 1 mm over the metasurface, when the spoof SPPs propagate into the surrounding medium at 10.2 GHz. The background medium here is air, and EM waves scattering into the air are focused.

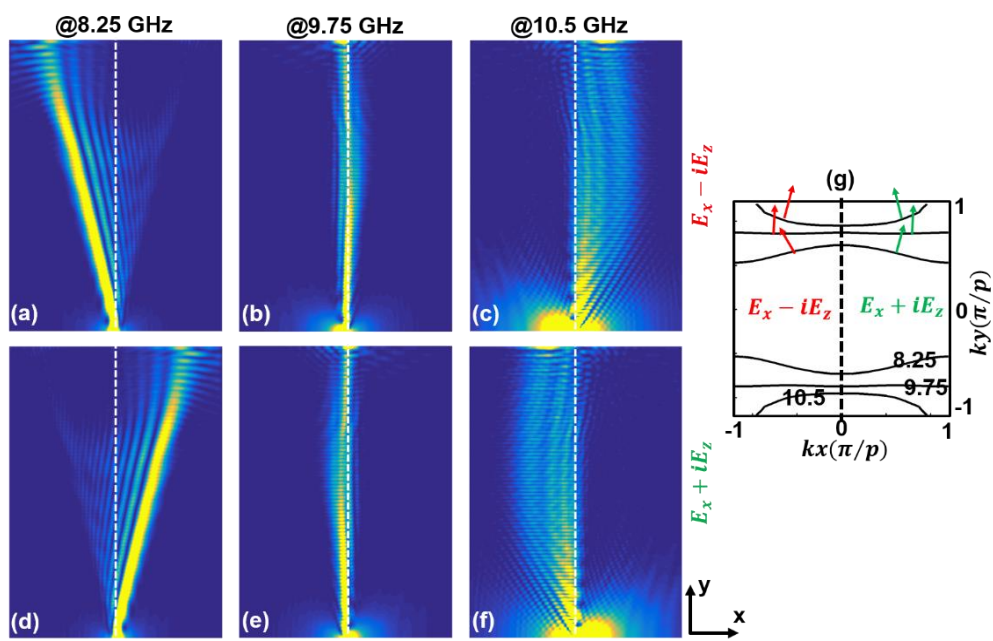

**Figure 6. Dispersion-dependent spin-momentum locking of spoof SPPs**. (a)-(c) Electric field intensity distribution of right-handed spin ($E_x - iE_z$) spoof SPPs at XY plane 5.6 mm over the metasurface for 8.25 GHz, 9.75 GHz, and 10.5 GHz, respectively. (d)-(f) Electric field intensity distribution of left-handed spin ($E_x + iE_z$) spoof SPPs propagates at XY plane 5.6 mm over the metasurface for 8.25 GHz, 9.75 GHz, and 10.5 GHz, respectively. (g) EFCs at

8.25 GHz, 9.75 GHz, and 10.5 GHz. Red and green arrows represent the propagation direction of right-handed spin and Left-handed spin, respectively.

The CHRM also shows strong abilities to control the transverse spins of spoof SPPs. Very recently, Konstantin Y. Bliokh *et al.* theoretically proved that the free-space light exhibits an intrinsic quantum spin Hall effect and the surface modes, such as SPPs and spoof SPPs, are with strong spin-momentum locking[23]. Therefore, the surface waves propagate along left and right will be with different transverse spins at the metal-dielectric interface, which is the so-called spin Hall effect and has been demonstrated in numerous experiments[24,25]. In our case, we put a dipole on the metasurface which will excite spoof SPPs with different transverse spins propagating left and right. By tailoring the dispersion of the metasurface, we can surprisingly control the transverse spins of the spoof SPPs. When the EFC is elliptical, the spoof SPPs propagating along left / right direction will be with right-handed spin ($E_x - iE_z$)/ left-handed spin ($E_x + iE_z$) as the normal case (Fig 5(a) and (d)). When the EFC is flat, the spoof SPPs only propagate along the y direction, and the right-handed and left-hand spin on the XZ plane will be canceled each other (Fig. 5(b) and (e)). When the EFC is hyperbolic, the spoof SPPs will carry opposite transverse spins (Fig. 5(c) and (f)) compared with the normal case. Moreover, with the angle-dependent local density of electromagnetic state, the field distribution of spoof SPPs with the spin-momentum locking are strongly directional, as shown in Fig. 6(a) and (d). This phenomenon physically arises from the spin-orbit coupling and can be explained by the dispersion of the CHRM as shown in Fig. 5(g). The spoof SPPs mode with right-handed/ left-handed transverse spins are in the left/ right wave vector space, however, the propagation direction of the modes should be perpendicular to the EFCs. As a consequence, the spoof SPPs will show anomalous transverse spins when the EFC is hyperbolic. Moreover, when the EFC is flat, the spoof SPPs can only propagate along the y direction, the spin of spoof SPPs on *XZ* plane will vanish[26].

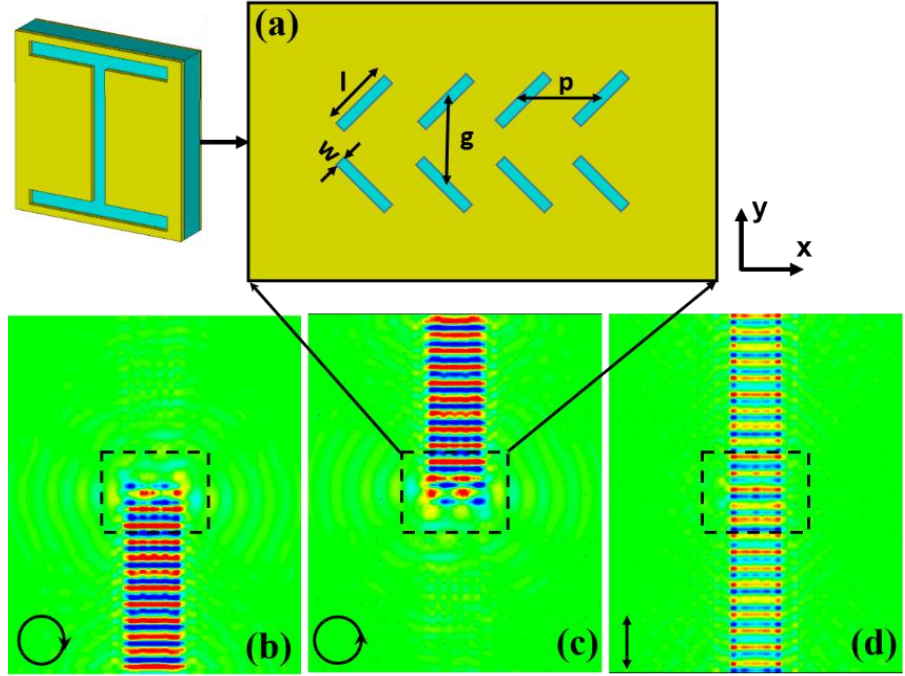

**Figure 7. Dispersion-dependent directional coupler.** (a) Scheme of the directional coupler. It is composed of two columns of sub-wavelength narrow apertures in the bottom metal film of the CHRM, where *w*=0.5 mm, *l*=3 mm, *p*=4 mm, *g*=*4.1* mm. (b)-(d) Simulated $E_z$ field distributions of soliton-like spoof SPPs on *XY* plane 5 mm over the CHRM when a plane EM wave with right circular polarization (b), left circular polarization (c), or linear polarization (d) is incident onto the metasurface from the bottom layer side at 9.75 GHz.

Based on dispersion-dependent spin Hall effect, we designed a coupler to launch diverge, soliton-like or convergent spoof SPPs directionally on the CHRM, controlled by the polarization of the incident EM waves. As shown in Fig. 7(a), the coupler is composed of two columns of sub-wavelength narrow apertures in the bottom metal film of the CHRM. It is well-known that if the coupler is well predesigned, destructive interference will occur at one side of the columns and constructive interference at the other side of the columns simultaneously, leading to the unidirectional launching of the surface wave[27,28] or spoof SPPs in our case. Here, by choosing *w*=0.5 mm, *l*=3 mm, *p*=4 mm, and $g=\lambda_0/4$=4.1 mm, where $\lambda_0$ is the wavelength of the spoof SPPs at 9.75 GHz, we can excite soliton-like spoof SPPs on the CHRM uni-directionally (Fig. 7(b)-(c)) or bi-directionally (Fig. 7(d)) at 9.75 GHz. By further engineering the apertures, we can even tailor the wavefronts of spoof SPPs with arbitrary shape[28]. The CHRM can work as a platform to study the properties of transverse spins of spoof SPPs with different dispersions and will find broad applications the signal processing.

## Conclusions

In this paper, we proposed and experimentally demonstrated an ultrathin MTT for manipulating spoof SPPs at low frequency. Comparing with three-dimensional metamaterials, this metasurface with 2D nature can achieve various dispersions, including elliptical dispersion, extreme anisotropic dispersion and hyperbolic dispersion with lower loss, the convenience of fabrication, and compatibility with the photonic integrated circuit. We demonstrated plenty of interesting phenomena based on this metasurface, including frequency-dependent spatial localization, non-diffraction propagation, negative refraction, and dispersion-dependent spin Hall effect, *etc.* Our metasurface will find broad applications in spatial multiplexers, focusing and imaging devices, planar hyperlens, dispersion-dependent directional coupler, and photonic integrated circuits. If involved with active components[29], such as semiconductor devices, we can even dynamically tailor the dispersion of the metasurface and control the manner of spoof SPPs, including propagation and spins. Besides, this anisotropic metasurface can also work as a brick of two-dimensional transformation optics based devices.

## Acknowledgements

This work was sponsored by the National Natural Science Foundation of China under Grants No. 61322501 and No. 61275183, the National Program for Special Support of Top-Notch Young Professionals, the Program for New Century Excellent Talents (NCET-12-0489) in University, the Fundamental Research Funds for the Central Universities, and the Innovation Joint Research Center for Cyber-Physical-Society Systems. Work at Ames Laboratory was partially supported by the U.S. Department of Energy, Office of Basic Energy U. Science, Division of Materials Sciences and Engineering (Ames Lab-oratory is operated for the S. Department of Energy by Iowa State University under Contract No. DE-AC02-07CH11358).